%% file: ms.tex
\documentstyle[11pt,preprint]{aastex}

\newcommand{\vhbb}{$\Delta V_{\rm HB}^{\rm bump}\,$}
\newcommand{\vhbbstar}{$\Delta V_{\rm HB}^{\rm \ast,bump}\,$}

\shorttitle{On the RGB bump in dwarf galaxies}
\shortauthors{Monelli et al.}

\begin{document}

\title{The ACS LCID project IV: detection of the RGB
bump \\ in isolated galaxies of the Local Group\altaffilmark{1}}

\author{M. Monelli\altaffilmark{2,3},
S. Cassisi\altaffilmark{4},
E. J. Bernard\altaffilmark{5},
S. L Hidalgo\altaffilmark{2,3},
A. Aparicio\altaffilmark{2,3},
C. Gallart\altaffilmark{2,3},
E. D. Skillman\altaffilmark{6}
}

\altaffiltext{1}{Based on observations made with the NASA/ESA {\it Hubble Space
   Telescope}, obtained at the Space Telescope Science Institute, which is
    operated by the Association of Universities for Research in Astronomy,
    Inc., under NASA contract NAS5-26555. These observations are associated
    with program 10505.}
\altaffiltext{2}{Instituto de Astrof\'{i}sica de Canarias, La Laguna, Tenerife,
    Spain; monelli@iac.es, carme@iac.es, antapaj@iac.es, shidalgo@iac.es.}
\altaffiltext{3}{Departamento de Astrof\'{i}sica, Universidad de La Laguna, 
    Tenerife, Spain}
\altaffiltext{4}{INAF-Osservatorio Astronomico di Collurania,
    Teramo, Italy; cassisi@oa-teramo.inaf.it}
\altaffiltext{5}{Institute for Astronomy, University of Edinburgh, Royal 
    Observatory, Blackford Hill, Edinburgh EH9 3HJ, UK; ejb@roe.ac.uk}
\altaffiltext{6}{Department of Astronomy, University of Minnesota,
    Minneapolis, USA; skillman@astro.umn.edu.}

\begin{abstract}
We report the detection and analysis of the red giant branch luminosity function bump
in a sample of isolated dwarf galaxies in the Local Group. 
We have designed a new analysis approach comparing the observed 
color-magnitude diagrams with theoretical best-fit color-magnitude diagrams
derived from precise estimates of the star formation histories of each galaxy. 
This analysis is based on studying the difference between the V-magnitude of 
the RGB bump and the horizontal branch at the level of the RR Lyrae instability 
strip (\vhbb) and we discuss here a technique for reliably measuring
this quantity in complex stellar systems.
By using this approach, we find that the difference between the observed and 
predicted values of \vhbb is $+0.13 \pm 0.14$ mag. This is 
smaller, by about a factor of two, than the 
well-known discrepancy between theory and observation at low metallicity 
commonly derived for Galactic 
globular clusters. This result is confirmed by a comparison 
between the adopted theoretical framework and empirical estimates of the \vhbb 
parameter for both a large database of Galactic globular clusters and for four other 
dSph galaxies for which this estimate is available in the literature. 
We also investigate the strength of the red giant branch bump feature ($R_{bump}$),
and find very good agreement between the observed and theoretically predicted 
$R_{bump}$ values.  This agreement supports the reliability of the evolutionary 
lifetimes predicted by theoretical models of the evolution of low-mass stars.
 
\end{abstract}

\keywords{
  Local Group ---
  galaxies: individual (Cetus dSph, Tucana dSph, IC1613 dIrr, Leo~A dIrr, LGS~3 dIrr/dSph) ---
  galaxies: stellar content ---
  stars: general ---
  stars: evolution}

\section{Introduction} \label{sec:intro}

In low-mass stars, the evolution of the H-burning shell along the Red 
Giant Branch (RGB) involves an extremely small mass ($\approx$0.001$M_\odot$ 
or less near the RGB Tip). Therefore, its crossing of any discontinuity in 
the H-abundance profile causes a change in the H-burning efficiency and,
in turn, a sudden change in the stellar surface luminosity. During the 
RGB evolution, the H-burning shell crosses the chemical discontinuity left 
by the convective envelope after the first dredge-up. When the shell 
encounters this discontinuity, the matter in the shell expands and 
cools slightly, causing a sharp drop in the stellar surface luminosity. 
When thermal equilibrium is restored, the stellar 
luminosity starts to monotonically increase up to the RGB Tip. As a 
consequence, the star crosses three times a narrow luminosity interval. 
This produces the characteristic bump in the theoretical 
luminosity function (LF). 
 
Since its detection in 47 Tucanae \citep{king85}, the RGB bump has become the 
crossroad of several theoretical and observational investigations \citep{fusipecci90,
ferraro90, cassisi97, alves99, zoccali99, bono01, riello03, dicecco10}. In order to 
detect the bump, a large sample of RGB stars is required -- $\sim10^3$ 
stars in the upper four magnitudes of the RGB according to \citet{fusipecci90} 
-- and until a few years ago it was detected only in metal-rich globular 
clusters (GC). In fact, with increasing cluster metallicity, the luminosity 
extension of the bump is larger and the bump is also shifted to lower luminosity,
where the RGB is more densely populated; both effects work in the direction of making 
the bump detection easier. However, in the last decade, the RGB bump has been detected 
in metal-poor GCs, due to the HST capability of imaging 
the dense GC cores and the availability of wide field-of-view CCD detectors 
on ground-based telescopes (\citealt{riello03}, see also \citealt{dicecco10}).
 
The parameter widely adopted to study the RGB bump brightness is the 
$\Delta V_{\rm HB}^{\rm Bump}= V_{Bump}-V_{HB}$ (\vhbb), that is, the V-magnitude 
difference between the RGB bump and the horizontal branch (HB) at the RR Lyrae 
instability strip level \citep{fusipecci90, cassisi97}. From an observational 
point of view, this has the advantage of being independent of distance and reddening,
but it requires that the instability strip is well populated \citep{ferraro99}. 
From a theoretical point of view, 
\vhbb depends not only on the bump level predicted by models, but also on 
the HB luminosity set by the value of the He core mass
at the He-flash. For a detailed discussion of the model uncertainties and 
the impact on the RGB bump properties see \citet{cassisi97}, \citet{salaris02}
and \citet{bjork06}.

The first exhaustive comparison between theory and observations of \vhbb in 
Galactic GCs was performed by \citet{fusipecci90}. Using the theoretical 
models available at that time, they found that the observed dependence of 
\vhbb on cluster metallicity was in good agreement with theoretical predictions, 
but theoretical \vhbb values were smaller than the observed ones by $\approx 0.4$ 
mag. For a long time, this result has been considered a clear drawback of 
canonical theoretical models of low-mass RGB stars. Therefore, it was suggested 
by \citet{alongi91} that this discrepancy could be removed by accounting for 
the efficiency of overshooting from the base of the formal boundary of the 
convective envelope, which shifts the bump level to a lower brightness 
because of the resulting deeper convective envelope. 
 
\citet{cassisi97} reanalyzed the problem using their updated 
canonical stellar models applied to a sample of 8 clusters with spectroscopic
determinations of [Fe/H] and [$\alpha$/Fe],  accounting for 
the effect of the $\alpha$-element overabundance. They concluded that their 
models provided a good match to the available observational data. 
\citet{zoccali99} and \citet{ferraro99} have produced larger observational 
\vhbb databases and compared these empirical data with suitable stellar models. 
From this comparison, one can draw the conclusion that lingering uncertainties 
on the HB theoretical brightness and the GC metallicity scale still leave open
the possibility that a discrepancy at the level of $\sim$0.20 mag between theory and 
observations may exist. The recent analysis made by 
\citet{dicecco10} with up-to-date theoretical models seems to support this.
For a detailed discussion about the differences between the stellar models adopted by
\citet{cassisi97} and the most recent theoretical framework, see 
\citet{dicecco10}.

All the empirical evidence quoted above refers to RGB bump detections in Galactic GCs. 
However, there is a growing number of 
detections of---single and double---RGB bumps in dwarf spheroidal (dSph) satellites of
the Milky Way (Sculptor: \citealt{majewski99}; Sextans: \citealt{bellazzini01}; 
Sagittarius: \citealt{monaco02}; Ursa Minor: \citealt{bellazzini02}; Leo~II: 
\citealt{bellazzini05}) while the first detection in an isolated Local Group 
(LG) galaxy was presented for DDO210 by \citet{mcconnachie06}. 

It is now well known that dSph galaxies are not simple stellar populations,
but rather they are a mix of populations, with different ages and
chemical properties. Therefore, it is clear that dSph galaxies are not the 
best target - when compared with \lq{simple}\rq\ Galactic GCs - in order to 
check the capability of the evolutionary framework to reproduce the RGB bump 
brightness. This notwithstanding, the RGB bump detections in dSph galaxies 
have been used for constraining properties (e.g., possible differences in 
age and/or metallicity) of the various stellar populations present in the 
corresponding stellar systems (see references above).

In the present work, we present a different approach for the analysis 
of the RGB bump, in particular for dwarf galaxies. Recently, we 
have carried out a project with the aim of studying a sample of isolated 
LG dwarf galaxies. A summary of the goals and results of the LCID project
({\itshape Local Cosmology from Isolated Dwarfs} \footnotemark[7])
can be found in C.\ Gallart et al.\ (in prep.). Here it is worth recalling that the main 
aim of the LCID project was to derive precise star formation histories (SFHs) over 
the lifetime of the selected objects, in order to obtain a detailed description
of their evolution and to analyze their properties in a cosmological context.
This was possible due to the observational capabilities of the ACS
camera aboard the Hubble Space Telescope for the Local Group galaxies 
Cetus, Tucana, Leo~A, IC~1613, and LGS~3\footnotemark[8]. 
The ACS observations allowed us to obtain color-magnitude diagrams (CMDs) of 
unprecedented 
depth and quality for these objects. The analysis of these CMDs allowed us to detect 
the RGB bump(s) in all five galaxies observed with the ACS.

\footnotetext[7]{http://www.iac.es/project/LCID}
\footnotetext[8]{The Phoenix dwarf is the sixth galaxy in the LCID project, but was 
observed with the WFPC2, and the WFPC2 data did not allow a robust detection of the 
RGB bump. Therefore, this galaxy is not included in the present analysis.}

Preliminary results on this issue have been presented for the Cetus dSph 
by \citet{monellicetus}. In the present paper, we extend the analysis and
present our approach to compare the observed and theoretical values of \vhbb. 
The essence of our method is to take advantage of the derived SFHs,
which provide a reliable description of the ages, chemical properties, 
and counts of the stars in both the RGB and the HB stages. 
The SFHs are used to calculate a synthetic CMD, which is then 
compared with the observed CMD.

The outline of this paper is as follows: in the next section we discuss the 
method and provide the relevant data about both the observational and the 
theoretical framework; in \S \ref{sec:compa} we compare theoretical 
predictions based on the retrieved SFHs with the observational measurements. 
We close with some discussion and final remarks.

\section{Data reduction and analysis: the \vhbb parameter estimate} \label{sec:data}


\begin{figure*}
\centering
\includegraphics[width=14truecm, height=7truecm]{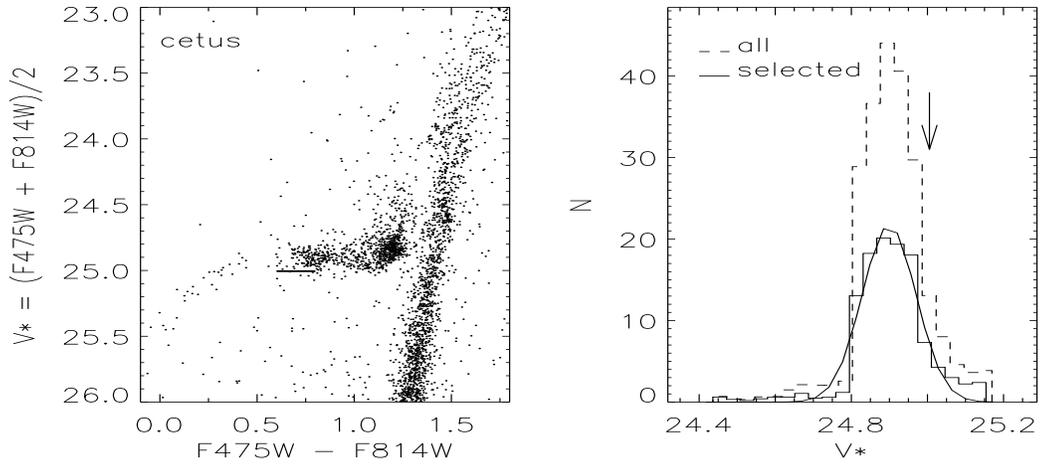}
\caption{This figure illustrates the technique adopted to estimate the magnitude
level of the HB. The left panel shows a zoom in the HB region of the observed CMD 
(in this example for the Cetus dSph). The horizontal line marks the HB level.
This is calculated fitting the magnitude histogram of the HB stars with a Gaussian
profile,
as shown in the right panel. This presents the histogram for all the stars
in the instability strip, and the adopted sub-sample in a color range 0.2 mag
wide. Note the similarity of the two histograms. In this particular case, the 
ZAHB level estimated using all the HB stars differs from the adopted value by 
0.01 mag.
\label{fig:hbobs}}
\end{figure*}


We present the analysis of the RGB bump for five of the LCID galaxies: Cetus,
Tucana, LGS~3, IC1613, and Leo~A.
An in-depth discussion of the data reduction and photometry can be found in 
\citet{monellicetus}. We note that the photometry was obtained 
using both the DAOPHOT/ALLFRAME \citep{alf} and DOLPHOT \citep{dolphot} 
packages. \citet{monellicetus} discuss the small zero points  
detected between the two photometry sets. However, note that the results of 
this paper, based on the magnitude difference of the RGB bump and the HB, 
are not affected by such small offsets.
The CMDs and SFHs of the five galaxies are being presented in a series of papers 
(Cetus: \citealt{monellicetus}; Tucana: M.\ Monelli et al., in prep.; IC1613: 
E.\ Skillman et al., in prep.; LGS~3: \citealt{hidalgolgs3}; Leo~A: \citealt{cole07}). 
The variable stars, and in particular the RR Lyrae used in this work, have been 
presented in \citet[][Cetus and Tucana]{bernard09a}, \citet[][IC~1613]{bernard10}, 
and \citet[][LGS~3, Leo~A]{bernard09b}.

Since galaxies are complex stellar systems with mixtures of stellar 
populations of different ages and metallicities, we devised a new approach 
to compare the theoretical and the observed \vhbb. In particular, 
we take advantage of the detailed knowledge of the SFHs of the sample galaxies 
to build a best-fit model, or {\itshape best fit} CMD\footnotemark[9], which 
was used to estimate the theoretical \vhbb to be compared with the observed one.
We emphasize that the SFHs and the best fit CMDs adopted in present analysis 
have been derived using the BaSTI stellar model library \citep{pietrinferni04}.


\begin{figure*}
\epsscale{1.0}
\plotone{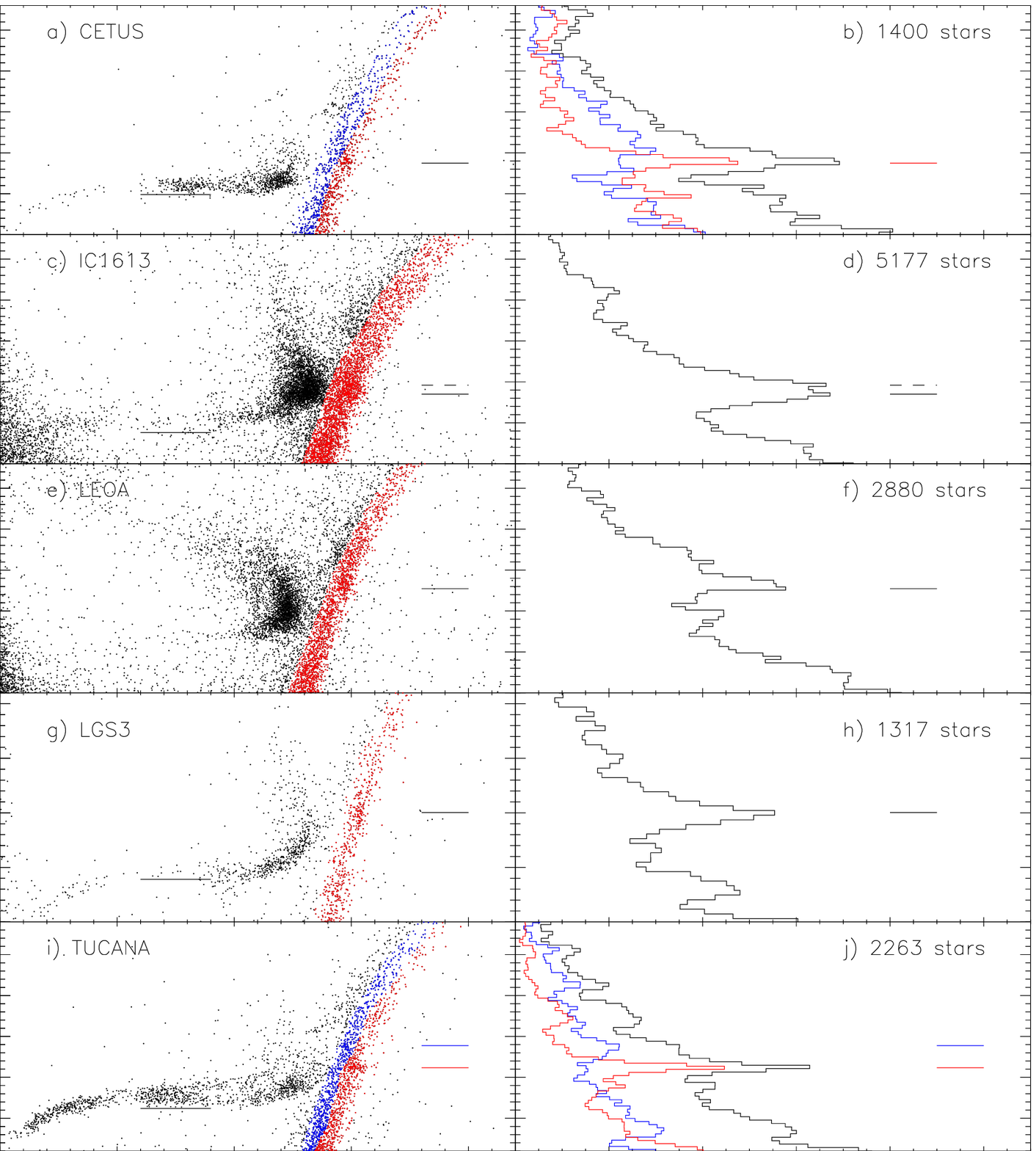}
\vspace{1cm}
\caption{{\sl Left panels}: The observed CMDs for the five dSph galaxies 
in our sample. The horizontal line marks the HB luminosity level.
{\sl Right panels}: The observed differential RGB LFs for the 
galaxies in our sample. In all panels the horizontal line marks the position of the 
detected RGB bumps. In the cases of Cetus and Tucana, the different colors refer to 
a star sample selection, adopted in order to trace in more detail the contribution 
of the various stellar populations to the the global LF (see the text for more details).
\label{fig:bumpobs}}
\end{figure*}


In the following, we outline the details of the method to estimate the \vhbb 
parameters in the empirical and in the {\itshape best fit} CMDs, 
presenting both the HB luminosity level estimates (\S \ref{sec:hb})
and the bump detections (\S \ref{sec:bump}).

Since the data collected for this project have been observed in the ACS 
$F475W$ and $F814W$ passbands, we define the $M_{V^\ast} = 
(M_{F475W} + M_{F814W})/2$ color, which is adopted 
for both CMDs\footnotemark[10]. The reason for this choice 
is twofold: {\em i)} the $M_{V^\ast}$ magnitude is quite close to
the Johnson V band commonly used to study the RGB bump, and {\em ii)} 
the $M_{V^\ast}$ luminosity of the HB is roughly constant as a function 
of color within the RR Lyrae instability strip (see below).

\footnotetext[9]{The best fit CMD was built using the IAC-star code \citet{iacstar},
available at the web page {\itshape http://iac-star.iac.es}}
\footnotetext[10]{With the usual notation, we will refer to $M_{V^\ast}$
and $V^\ast$ for the absolute and apparent magnitude, respectively.}

\subsection{Estimate of the HB luminosity} \label{sec:hb}

In concordance with other studies, we estimate the HB luminosity  at the
color corresponding to the RR Lyrae star instability strip ($0.4 < 
M_{F475W} - M_{F814W} < 0.9$ mag). However, we estimate that the mean
magnitude  of the HB changes by $\Delta M_{V^\ast} \approx 0.05$ mag from the
bluest to the  reddest limit of the instability strip. Therefore, we obtained
the HB luminosity  in a 0.2 mag wide color bin consistent with the center of
the instability strip,  centered at $M_{F475W} - M_{F814W} = 0.65$ mag. Note
that the change in the  HB luminosity is $\leq$ 0.03 mag when using the full
color range instead of the adopted range which is well within our quoted
uncertainties.

In the case of the observed CMD, we cannot estimate the Zero Age
Horizontal Branch (ZAHB) by adopting the lower envelope of the observed
stellar distribution (as is usually done in the case of Galactic GCs) since
the presence of blends, unresolved galaxies, and/or RR Lyrae variables  with
poorly measured mean magnitudes can affect the estimate of the
\lq{real}\rq\ ZAHB level. Therefore, we have adopted a different approach. By
using the synthetic, best fit, CMDs, we have verified that the luminosity
level at 1.5$\sigma$ fainter than the peak magnitude of the HB LF, calculated by
means of a Gaussian fit, nicely reproduces the location of the theoretical HB
lower envelope for all of the galaxies in our sample; the same method was then used
for estimating the ZAHB level in all galaxies.

Figure \ref{fig:hbobs} shows the example of Cetus. The left panel shows a zoom
of the CMD, where the thick black line marks the estimated ZAHB. The right panel
shows the histogram of the HB stellar magnitudes. The dashed line represents all
the HB stars in the color range 0.4 $< M_{F475W} - M_{F814W} < $0.9, while the
solid line represent the adopted sub-sample in a restricted color range.
Over-plotted on this we present the fitted Gaussian profile and the position of the
estimated ZAHB, marked by the arrow. The observed  HB levels for all the galaxies
of our sample are shown in Figure \ref{fig:bumpobs},  left panels, and summarized
in Table \ref{tab:tab1}. 


\input{tab01}


In the case of the best fit CMD, the HB is populated by stars which are,
by construction, on the ZAHB, or eventually brighter due to evolutionary 
effects (which are modeled in IAC-star, \citealt{iacstar}). 
Therefore, we can safely adopt the minimum of the distribution
to characterize the HB luminosity. We adopted the same color range as for 
the observed CMD. The derived values are presented in Table \ref{tab:tab2}.


\begin{figure}
\epsscale{1.0}
\plotone{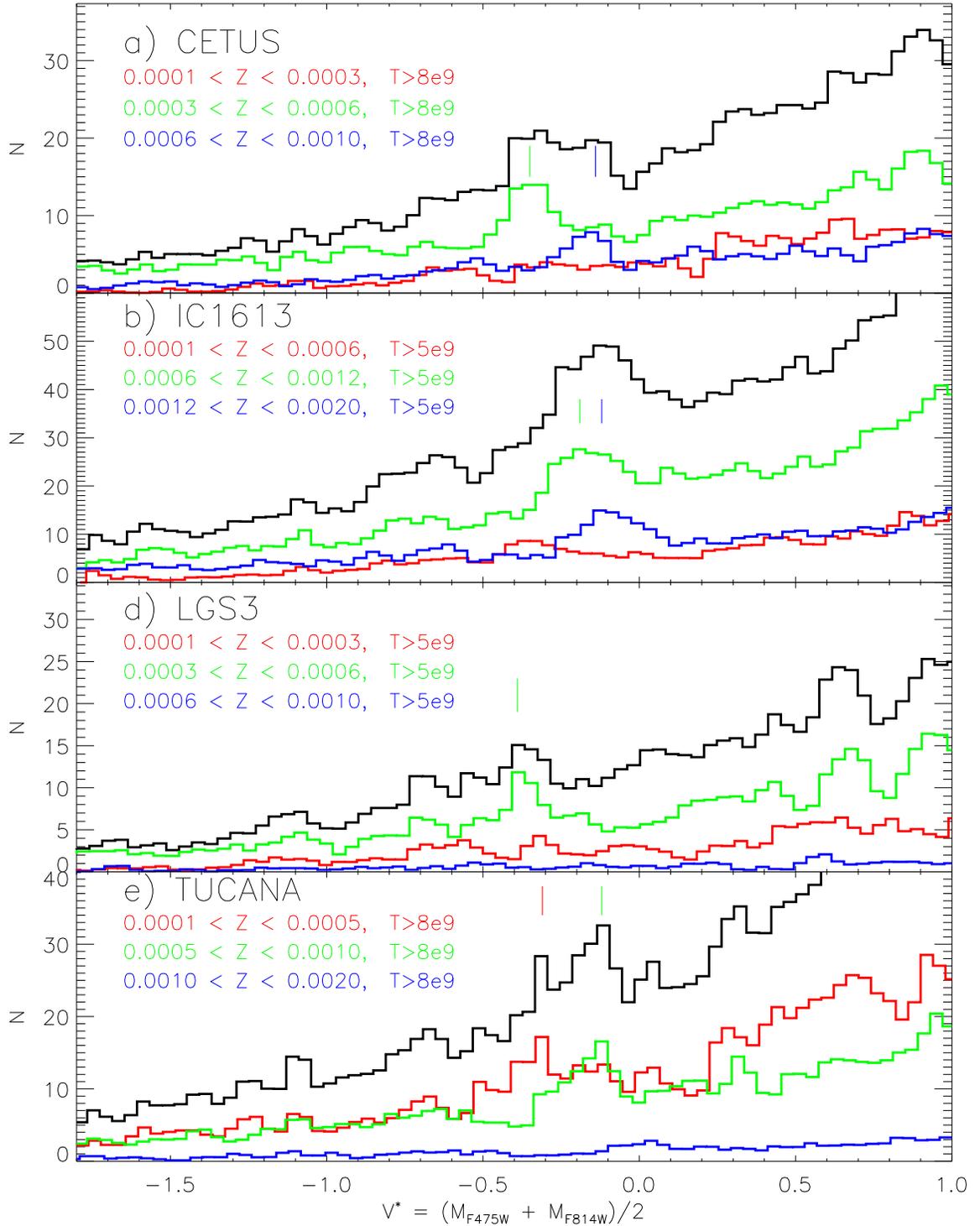}
\vspace{0.5cm}
\caption{The four panels show the histograms of the RGB stars of the best fit 
CMDs for the four galaxies. The black lines represent the global histogram, 
while the lines in color show the histogram of the sub-populations with 
labeled ages and metallicity. With this approach we were able to constrain 
which population is the main contributor to the RGB bump. Note that the 
metallicity ranges change from one galaxy to another, due to the different
details of the individual SFHs.
\label{fig:histo}}
\end{figure}


\subsection{Detection and characterization of the RGB bump} \label{sec:bump}

The RGB bump appears as an obvious feature in the five observed CMDs 
(see Figure \ref{fig:bumpobs}). To estimate the magnitude of the 
observed bumps, we first performed a selection of the RGB regions, 
in order to avoid the contamination by the stars in different evolutionary 
stages (colored dots in Figure~\ref{fig:bumpobs}, left panels). We then 
calculated the luminosity function of the RGB using a running average, 
to smooth out the binning effect. The dimension of the smallest bin is 
0.04 mag, which is the assumed error on the bump magnitude estimate.
Figure~\ref{fig:bumpobs} shows the location of the detected 
RGB bump for the five galaxies, both in the CMD and in the differential LF. 
Note that the histograms in the right panels have been arbitrarily normalized
for clarity.


\input{tab02}


The same approach was applied to the best fit CMD. However, the information 
from the SFH, reflected in the best fit CMD, allows a deeper analysis of 
this diagram. Since these galaxies have composite stellar populations, 
their RGB bumps can be the result of the superposition/contribution 
of different sub-populations with different ages and/or chemical compositions. 
Therefore, using the information from the SFHs, we can identify which stellar 
component is the dominant contributor to the RGB bump. To do this, we tested 
different metallicity bins, building the corresponding LFs and comparing them
with the global one. A few tests allowed us to find out the optimal age and 
metallicity ranges for each galaxy. The final result of this analysis is 
shown in Figure~\ref{fig:histo}, for the four galaxies for which we have been able 
to estimate both the HB and RGB bump luminosity level. Note that the age 
and metallicity bins slightly change from one object to another, because 
of their different evolutionary histories, and that we only adopt a weak
constraint on the age, since the dominant parameter affecting the RGB bump 
brightness is the metallicity.

Table \ref{tab:tab2} summarizes the derived quantities for the best fit CMDs:
the estimated luminosity of the HB and RGB bump, and the 
calculated \vhbbstar, together with the adopted age and metallicity 
ranges. The error associated to the RGB bump brightness has been estimated
as half of the width of the magnitude bin of the RGB luminosity function.

We wish to emphasize that, since in retrieving the SFH we have not taken into
account the RGB and HB stages \citep{monellicetus, hidalgolgs3}, the obtained 
model predictions are the results of a \lq{blind}\rq\ technique for which no 
{\sl a priori} constraints (such as star counts and CMD location)
have been applied.
Consequently, the comparison 
between theory and observations can provide useful (albeit indirect)
clues concerning the accuracy of the theoretical framework reproducing the 
RGB bump brightness in composite stellar populations. 

\subsection{Notes on individual galaxies} \label{sec:notes}

{\bf TUCANA} -- A visual inspection of the observed CMD reveals the presence of two
clear over-densities along the RGB: the brightest is located on the blue edge
of the RGB, while the faintest is on the red edge. To further investigate 
the possible occurrence of two distinct bumps in Tucana, we calculated the 
LF of both the red and the blue portion of the RGB, as shown in panel 
$(i)$ of Figure \ref{fig:bumpobs}. The corresponding histograms (in color in 
panel $j$), shown together with the global LF (solid black line), support 
the detection of two distinct RGB bumps in Tucana, separated by 0.27 $\pm$ 0.06 mag.

The presence of a double bump has direct connections with the SFH of Tucana.
Two of the main results of the LCID project concerning this galaxy are: \\
$i)$ Tucana experienced a strong initial burst of star formation, at the oldest
possible ages ($>$ 12.5 Gyr ago, implying a substantial population of metal
poor stars), and followed by a steady decrease until reaching a complete 
termination roughly 8 Gyr ago (Monelli et al. in prep.) \\ 
$ii)$ Tucana hosts two distinct populations of RR Lyrae variable stars, characterized
by different luminosity, pulsational properties, and radial distributions
\citep{bernard08}. \\
These two findings suggest that the Tucana dSph was able to form two different
generations of stars with different chemical properties in a very short time
interval early on, so both populations are old enough to produce RR Lyrae. 
Therefore, there is a good qualitative agreement between these results
and the independent detection of two RGB bumps, which can be interpreted
using the LF of the best fit CMD. In fact, the latter also possesses a similar 
double peak (Figure \ref{fig:histo}, panel $d$).
Interestingly, the two peaks are associated to two populations of different 
metallicity: the brighter is made of stars with metallicity in the most 
metal-poor range, 0.0001 $< Z < $ 0.0005, while the fainter is populated 
by stars in the range 0.0005 $< Z < $ 0.001. The magnitude difference between 
these two features, 0.19 $\pm 0.06$ mag, is in good agreement with the magnitude 
difference of the observed bumps, 0.27 $\pm 0.06$ mag. Moreover, we verified that, as 
expected, the most metal-poor stars populate the blue side of the RGB, while
the more metal-rich are located on the red side. Therefore, as for the observed
bump, the brightest bump is also the bluest, and the faintest is also the reddest.
Another interesting aspect is that the two Tucana bumps are the narrowest 
among the LFs of the five galaxies, with estimated dispersions of $\sigma$ = 
0.04$\pm0.01$ mag and 0.06$\pm0.01$ mag for the faintest and the brightest, 
respectively. This is possibly linked to the intrinsic short duration
of the bursts of star formation (Monelli et al. in prep.). 

{\bf CETUS} -- The RGB bump of Cetus is also very prominent, as shown by Figure 
\ref{fig:bumpobs}, panel $(a)$. Interestingly,
the position of the bump is similar to the faintest bump detected in Tucana, 
and both are located on the red edge of the RGB. The bump clearly stands out in the
observed LF as well (Figure \ref{fig:bumpobs}, panel $b$). The analysis of the
red and the blue portion of the RGB (red and blue lines) shows that the global 
peak (solid black line) seems to be the result of the superposition of two 
different features. The main narrow peak is populated by stars on the red side 
of the RGB, while the bluer portion mostly contributes to broaden it. The width 
of the main peak, $0.06\pm0.01$ mag, is slightly larger than the corresponding 
faint bump in Tucana.

The comparison of the SFHs of Cetus and Tucana, presented in Monelli et al. (in prep.), 
shows that Cetus did not experience an initial burst as strong or as 
early as in Tucana. In particular, Monelli et al.\ found that the main star formation
peak is $\sim$ 1 Gyr younger in Cetus than in Tucana. These differences, and 
the smoother SFH of Cetus, plausibly explain both the absence of a second brighter 
bump in Cetus, and the rather homogeneous properties of the RR Lyrae stars, which 
do not show any clear dichotomy, as in the case of Tucana \citep{bernard09a}.

The theoretical LF (Figure \ref{fig:histo}, panel $a$), shows quite a broad peak, 
formed by the contribution of two populations, in the ranges 0.0004 $<$ Z 
$<$ 0.0007, and 0.0007 $<$ Z $<$ 0.001. The LF of the most metal-poor stars 
does not show any clear feature.

{\bf IC~1613} -- The bump region in the LF of IC~1613 is the widest among
our sample ($\sigma $= 0.15 $\pm$ 0.02 mag). This can be explained by
the extended SFH of this galaxy, which was continuous from the oldest 
epoch to the present time. There are hints of the presence of two peaks, 
separated by $\approx$ 0.12 mag. The theoretical LF (Figure \ref{fig:histo}, panel 
$d$) shows that two populations are contributing to the main peak, in a wide 
range of metallicities. The corresponding peaks are separated by $\approx$ 
0.07 mag.

{\bf LEO~A} -- This galaxy has the most unusual SFH in our sample. \citet{cole07}
found very low level of star formation starting $\approx$13 Gyr ago, in agreement
with the small number of RR Lyrae stars detected (8: \citealt{dolphin02}, and 10:
\citealt{bernard09b}). \citet{cole07} further found that 90\% of all star formation
occurred in the last 8 Gyr, resulting in the only galaxy in the LCID sample with a
dominant intermediate-age population. As a result, the HB inside and to the blue
of the instability strip is very scarcely populated by old stars. Moreover, the
time sampling in our ACS observations was poor and not adequate to accurately
estimate the periods, and therefore the mean magnitudes, of the RR~Lyrae stars.
For these reasons, Leo~A is not included in our analysis.


\input{tab03}


\section{Comparison between theory and observations} \label{sec:compa}

In the previous section, we have discussed our approach to attribute 
the contributions of the various sub-populations present in the galaxy
to the RGB bump feature(s) in the global LF. When a double RGB bump is 
present in the empirical CMD, we checked the theoretical LF to determine 
which stellar components are responsible for the observed feature. 


\begin{figure}
\epsscale{1.20} 
\includegraphics[width=9truecm, height=8truecm]{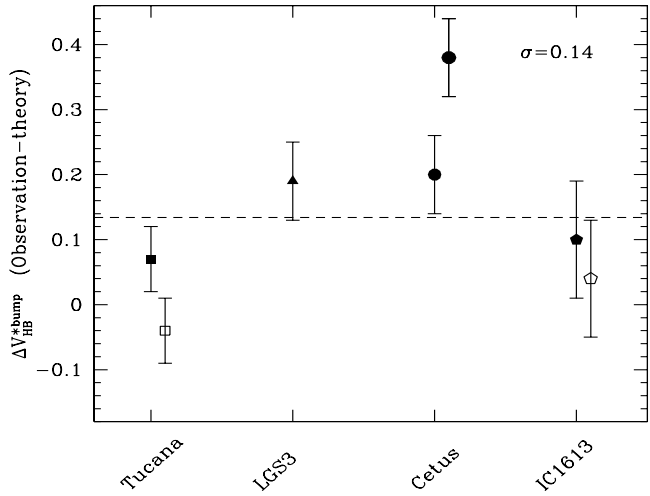} 
\caption{The difference between the \vhbbstar parameter as
estimated from the observed CMDs and the synthetic ones (see text for more
details). In the cases in which a double RGB bump has been observationally
detected, the full square points correspond to the fainter RGB bump detection. In
case of Cetus, since the model fit has two distinct RGB bumps,
we show the magnitude difference with respect the observational detection for
both the theoretical RGB bumps. The dashed horizontal line represents the mean 
value of the difference, while the standard deviation with respect to the mean is
labeled. 
\label{fig:teo-obs}} 
\end{figure}


Since the HB luminosity level is strongly dependent
on the metallicity, comparing theoretical predictions and observations of \vhbbstar
in stellar systems with mixed stellar populations with different metallicities
could be liable to uncertainties. However, by using our best fit CMDs, we have
tested that the more metal-rich populations are either too metal rich to 
populate the RR Lyrae instability strip, or that the difference in the HB 
luminosity level (the lower envelope of the two distributions within
the instability strip) with respect to the lower metallicity stellar population  
contributing to the RGB bump feature(s), is of the order of 0.02-
0.03 mag, therefore within the error bars of our estimate from the observed CMD.
Thus, when estimating the empirical \vhbbstar, we have
adopted for all galaxies, the magnitude corresponding to the lower envelope of 
the HB distribution, estimated as detailed above.

In Figure~\ref{fig:teo-obs}, we show the $V^\ast$ magnitude difference between the 
\vhbbstar parameter as estimated from the observed and the synthetic CMDs. The
error on this difference has been obtained by summing in quadrature the errors on
the observational and theoretical estimates. Note that the average
difference is $\sim+0.13 \pm 0.14$ mag. This result
appears in qualitative agreement with the one obtained by comparing theory with
observations of Galactic GCs concerning the \vhbb parameter: the observed
magnitude difference between the RGB bump and the HB appears lower than expected
on theoretical grounds. Nevertheless, the difference estimated for our sample of
dwarf galaxies appears significantly less than that obtained for Galactic GCs.


\begin{figure}
\epsscale{1.0}
\plotone{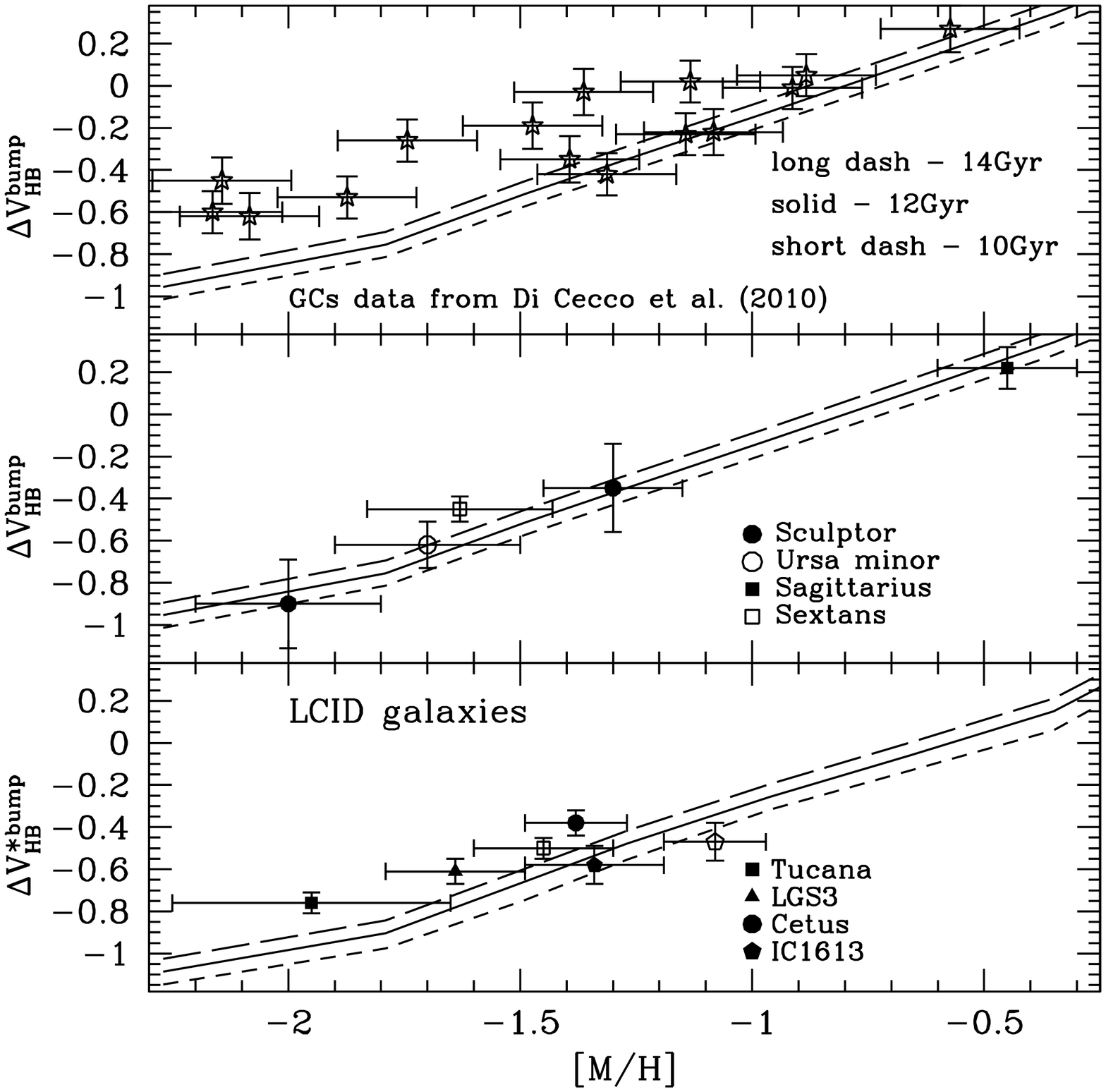}
\caption{{\sl Top panel}: The \vhbb parameter as a function of the metallicity 
for the sample of GCs studied by \citet{dicecco10}. The lines correspond to 
the theoretical predictions for ages of 10, 12, and 14 Gyrs (from bottom to top). 
{\sl Center panel}: as top panel, but for a selected sample of LG dwarf
galaxies for which estimates of the \vhbb parameter are available in the literature. 
The lines represent the same theoretical predictions shown in the top panel. 
{\sl Bottom panel}: as top panel, but in this case the estimates of the \vhbbstar 
parameter for the LCID galaxies are compared with the same theoretical predictions 
used in the other panels but transformed to the $V^\ast$ magnitudes. The same 
symbols (full and open) are used for the galaxies for which a double RGB bump has 
been detected. 
\label{fig:dwarfgc}}
\end{figure}


In order to allow a more direct analysis of how present results compare with 
those obtained for galactic GCs, we show in Figure~\ref{fig:dwarfgc} the 
comparison between theory and observations in the \vhbb -
metallicity diagram, for a large sample of Galactic GCs investigated by 
\citet{dicecco10}\footnotemark[11], for a selected sample of LG dwarfs for which 
\vhbb estimates are available in literature (see data in Table~3), and 
for the LCID dwarf sample\footnotemark[12]. In the latter case, the comparison
between theory and observations has been performed by using the \vhbbstar 
parameter, after transferring the model predictions into this observational plane. 

We note that the adopted theoretical
framework is the one provided by \citet{pietrinferni04}, i.e., the same adopted by
\citet{dicecco10}, and by our group in the framework of the LCID project for
retrieving the SFHs of the selected dwarfs. For the galaxies in our sample, we adopt
the metallicity distribution coming from the SFH analyses. That is, we have chosen
the metallicity value corresponding to the average of the metallicity range of
the stellar populations contributing to the RGB bump in the best fit CMD (see the
discussion in section \S \ref{sec:bump} and data in Table~2) and a spread equal to
the half width of the same interval\footnotemark[13].

\footnotetext[11]{We selected a subsample of 15 GCs with accurate ground based 
photometry, adopting the iron content from \citet{carretta09} and
an average $\alpha$-enhancement = 0.3 dex to calculate the global metallicity.}

\footnotetext[12]{ We note that, since we are comparing the \vhbbstar
parameter with the theoretical estimate for a single, simple stellar population,
for each RGB Bump detection in a given galaxy we should evaluate the observed \vhbbstar 
parameter by considering the HB luminosity level of the corresponding stellar population 
producing this feature. However, as discussed at the beginning of this section, this 
would cause a change in the \vhbbstar parameter of about 0.02-0.03 mag, i.e., small
with respect to the empirical uncertainty in the estimation of the HB luminosity level. 
Therefore, we have decided to not apply any correction to the \vhbbstar values listed 
in Table~1.} 

\footnotetext[13]{We wish to note that, at odds with the
case of Galactic GCs, this metallicity range is not related to the observational
uncertainty in the metal content of the stellar population, but it provides an
indication of the metallicity interval covered by the various sub-populations
contributing to the RGB bump feature in each dwarf in our sample.}

Since detailed SFH analysis, similar to those for our LCID sample, do not exist for the LG dwarfs
selected from the literature, we are not able to properly characterize the
particular stellar population(s) contributing to the RGB bump. Thus, we compare
theoretical predictions with the measured \vhbb by using the literature data at
their face value. Table \ref{tab:tab3} summarizes the \vhbb values from the
original papers, the adopted metallicities, and the related references. In
particular, the metallicity values have been taken, when available, from the most
recent spectroscopic measurements, or from the comparison with fiducial RGB loci of
selected Galactic GCs. We wish to note that the adopted empirical estimates for
the \vhbb parameter are not fully self-consistent due to the differences in the
adopted procedure for estimating the HB luminosity level and/or the approach used
for estimating the average metallicity of the stellar population to which the
\vhbb parameter is associated. Nevertheless, it is worth noting that this is, so
far, the largest data-set ever collected for the \vhbb parameter of LG dwarf
galaxies.

The comparison between the theoretical predictions and the empirical \vhbb (or
\vhbbstar) estimates show that, despite the quoted limitations of the observational data
for the dSphs, a better agreement exists with the theoretical predictions in the
metallicity range explored with respect to the similar comparisons for 
Galactic GCs.


\begin{figure}
\epsscale{1.20}
\includegraphics[width=9truecm, height=8truecm]{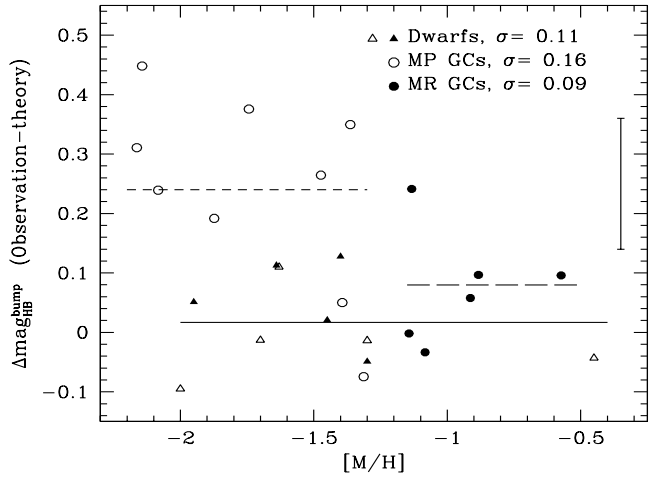}
\caption{The difference between theory and 
observations concerning the magnitude difference between the RGB bump 
and the HB for the Galactic GCs sample of \citet{dicecco10} 
(open circles: GCs more metal-poor than $[M/H]=-1.3$; full circles: GCs 
more metal-rich than $[M/H]\sim-1.1$) and the sample of dwarf 
galaxies accounted for in the present analysis (full triangles: LCID 
galaxies, open triangles: literature data). The short dashed line represents 
the mean value for the metal-poor GCs, the long dash line shows the mean value for 
the metal-rich GCs, while the solid line represents the same quantity but for the 
dwarfs galaxies in our sample. The standard deviation with respect to the 
mean is labeled in each case. The average error bar for both the GCs and 
the dSphs not included in the LCID sample is also shown; for the LCID galaxies the 
error bars are the same as in Figure~\ref{fig:teo-obs}.
\label{fig:teo-obs-tot}}
\end{figure}


To investigate further the level of agreement between theory
and observations for the various samples of objects, we have computed for each
object the difference between the empirical estimate of the \vhbb
parameter and the corresponding theoretical prediction for a mean age of
12.8~Gyr\footnotemark[14] and
the metallicity appropriate for the object under scrutiny. These differences
are plotted in Figure~\ref{fig:teo-obs-tot}. This figure shows
that for the more metal-poor GCs there is a mean difference
with respect to the theoretical predictions of about 0.24 $\pm$ 0.16 mag. 
For the more metal-rich GCs and for the dwarf
galaxies this average difference decreases to 0.08 and 0.02 mag respectively, with a
standard deviation of 0.09 and 0.11 mag respectively. In addition, note that for
the dwarfs there is no evident trend with the global metallicity. These results
support the conclusions based on the data in Figure~\ref{fig:teo-obs}, and seem to
point toward the presence of a problem in measuring the \vhbb parameter in
metal-poor GCs or, alternatively, to a problem in the GC metallicity scale. 
\footnotetext[14]{This age value has been adopted as the mean age of the Galactic
GC systems according to the assumption made by \citet{amarin09}. 
Note that a change of $\pm1$~Gyr would change the theoretical predictions
by about $\pm0.03$mag, which would have a marginal impact on our analysis.} 

An alternative possibility is that the complexity of the SFHs of the dwarf galaxies under investigation,
can contribute to \lq{mask}\rq\ the existence of a clear disagreement between
theory and observations. To clarify this point, we performed a test devoted to 
verify if - and eventually at what extent - the RGB bump brightness detection in the 
{\it global} stellar population would be affected when removing sub-populations 
characterized by a different spread in age and metallicity. It is clear that this 
analysis can be performed only by using the synthetic best fit CMDs. Specifically, 
we removed from the global CMD all the stars younger than 6.5 Gyr and 
progressively increased this age limit with bins of 1 Gyr, i.e., 
removing all the stars younger than 7.5, 8.5, ..., Gyr. In each case, we 
measured the brightness of the RGB bump of the remaining sample of stars.
The same test was performed subtracting stars older than a certain age (12.5 Gyr),
and decreasing this age limit (11.5, 10.5, ... Gyr). In both cases we 
found that, at a certain point of this procedure, the RGB bump disappears, as expected. 
However, until this moment, the magnitude difference between the RGB bump in the global 
and in the selected CMD is of the order of few hundredths of magnitude, well within
the uncertainties. We found the same effect in another analogous test where the stars
were removed according to their metallicity. 

As far as it concerns the HB luminosity level within the RR Lyrae instability strip, 
as well known it is not affected by the presence of an age spread in
the stellar population, and the impact of the existence of a metallicity spread has
been discussed - at least in the case of our sample of galaxies - in
section~\ref{sec:hb}. This analysis confirms that the complexity of the SFH - at
least in the case of our dwarfs galaxies - does not hamper the possibility of a
meaningful comparison between theory and observations. However, one has to bear in
mind that, in case of a more complex, or peculiar, SFH, this result might not be
valid; therefore it is important to investigate in detail each individual
case. 

For our purpose, it is enough to remark that the existence of a better
agreement between theory and observations with respect to Galactic GCs cannot be an
artifact due to complex stellar
populations. In order to understand the causes of the discrepancy existing for
Galactic GCs, a detailed analysis of the possible observational biases
(uncertainties) is mandatory, but this is beyond the scope of the present work.

\subsection{The $R_{bump}$ parameter}\label{sec:rbump}

In the previous section, we have investigated the level of agreement between the
observational measurements of the RGB bump brightness in our sample of dwarf
galaxies and the theoretical predictions as obtained both by synthetic CMDs,
reproducing the complex SFH of these galaxies, and by stellar models covering a
range of age and metallicity. However, there is another important piece of
information that can be retrieved from the RGB in the observed CMDs, i.e., the star
counts. It is well known \citep[for a review see][and references 
therein]{salaris02} that an appropriate comparison between the empirical and
theoretical RGB LFs is one of the best tools for checking the capability of
stellar models to predict the evolutionary lifetimes along this
important evolutionary stage.

In particular, \citet{bono01} defined a new observable, the $R_{bump}$, parameter,
defined as the ratio between the star counts across the RGB bump and in a region
located at a fainter magnitude along the RGB. They used this new parameter for
investigating the occurrence of non-canonical deep-mixing during the RGB bump stage
(see \citealt{weiss00}) in a large sample of Galactic GCs. As a result of their
analysis, they found a very good agreement - with very few exceptions - between
theory and observations.


\begin{figure}
\epsscale{1.20}
\includegraphics[width=9truecm, height=8truecm]{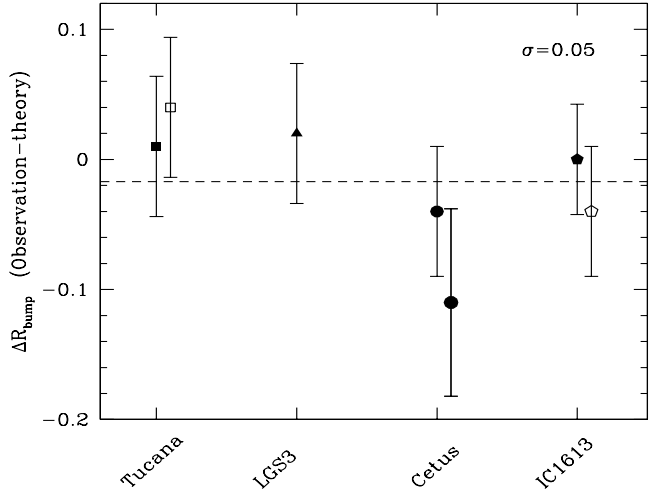}
\caption{The difference between the $R_{bump}$ parameter as estimated 
from the observed CMDs and the synthetic ones (see text for more details). 
As in Figure~\ref{fig:teo-obs}, since the best-fit model for Cetus shows 
two distinct RGB bumps, we show 
the $R_{bump}$ difference with respect to the observational detection for both 
the theoretical RGB bumps. The dashed horizontal line represents the mean 
value of the difference, while the standard deviation with respect to the 
mean is labeled.
\label{fig:rbump}}
\end{figure}


Due to the evident complexity of the SFH in dwarf galaxies with respect of Galactic
GCs\footnotemark[15], our CMDs are not the best tool
for investigating the occurrence of deep-mixing in RGB stars. This notwithstanding,
we have decided to measure the $R_{bump}$ in our empirical CMDs and compare these
measurements with the corresponding ones provided by the accurate best fit
synthetic CMDs. This comparison allows us to check if the predicted star counts
along the RGB are in agreement with the empirical data. In this context, we wish to
emphasize again that, when obtaining the best fit synthetic CMD for each galaxy in
our sample, we have not imposed any constraints on this evolutionary stage.
\footnotetext[15]{However, we refer the interested reader to \citet{cassisi08} for a
short review about the growing observational evidence that many - if not all -
Galactic GCs host multiple stellar populations.}

As far as it concerns the method for measuring the $R_{bump}$ parameter, we have
adopted the same method outlined by \cite{bono01}. The values measured in
the empirical and synthetic CMDs are provided with the corresponding
$1\sigma$ errors in Tables \ref{tab:tab1} and \ref{tab:tab2}. In
Figure~\ref{fig:rbump}, we plot the differences between the observed value(s) of
$R_{bump}$ in each dwarf in our sample, and the theoretical value(s) as estimated
from the synthetic best fit CMD. This figure shows 
very good agreement between theory and observation: the mean
difference is equal to $\sim -0.02 \pm 0.05$.
This result provides further evidence of the capability of retrieving the
SFH of these complex stellar systems as well as of the evolutionary theoretical
framework of providing correct predictions of stellar lifetimes along the RGB
stage.

\section{Summary and Conclusions} \label{sec:summa}

We have presented an homogeneous analysis of the RGB for five isolated dwarf galaxies in
the LG. This analysis allowed us to obtain the first detection of the RGB LF bump
in these stellar systems. In order to perform a direct comparison with theoretical
data, we measured the \vhbbstar parameter in four galaxies in our sample. 

We have designed and implemented a new approach to estimate this parameter, 
based on comparing an observed CMD with a theoretically calculated CMD.
This method takes advantage of the detailed knowledge of the stellar
content of the sample galaxies, in terms of age, metallicities, and star counts, to
identify which sub-population is actually contributing to form the RGB bump. This
is a fundamental advantage when working with complex stellar populations.

The comparison between the observations and the theoretical predictions reveal that
the well-known discrepancy between the observed and predicted brightness of the RGB
bump seen in Galactic GCs is smaller, by more than a factor of two, in our dSph
sample. This evidence is supported by the inclusion of four additional dSph, for
which the \vhbb values are available in the literature. 

It may be somewhat surprising that the level of agreement between theory and
observations is better for complex stellar populations (dSphs) than for
simple stellar populations (Galactic GCs). However, although the complex nature of
the stellar population could represent a drawback for clearly detecting the
observational features and reliably identifying the stellar component responsible
for the observed feature - and in this context our approach is a big advantage -,
the statistical significance of the stellar samples is quite larger. In the case
of our sample of dwarf galaxies, we have investigated if the complexity of the SFH
can affect our estimates of the \vhbb parameter in the empirical CMDs. As a result,
we have found that - at least for our sample of dSph - the measurement of the \vhbb
parameter is not significantly affected by the mix of sub-populations in the host
galaxy.

Our analysis has also shown the existence of a very good
agreement between observations and theoretical predictions concerning the
$R_{bump}$ parameter. This agreement represents supporting evidence that the adopted
theoretical framework is providing accurate estimates of stellar evolutionary
lifetimes.

The present analysis also emphasizes the importance of combining the detection and
characterization of the RGB bump(s), with other information concerning the
properties of the stellar populations hosted by a galaxy. In this context, it is
particularly important to note the case of the Tucana dSph; the detection
of a double bump can be correlated both with the properties of the RR Lyrae stars
and the SFH of this galaxy, supporting the coexistence of two old stellar
populations, with slightly different metallicities, formed during an initial strong
episode of star formation. This provides support for the use of the
RGB bump as a stellar population indicator for CMD which do not reach the
oldest main sequence turn-off.

\acknowledgments

{\it Facilities:} \facility{HST (ACS)}.

We warmly thank our referee for her/his pertinent comments and suggestions that have improved the
readability and the content of this paper.
Support for this work was provided by NASA through grant GO-10515 from the Space
Telescope Science Institute, which is operated by AURA, Inc., under NASA contract
NAS5-26555, the IAC (grant 310394), the Education and Science Ministry of Spain
(grants AYA2004-06343 and AYA2007-3E3507). This research has made use of NASA's
Astrophysics Data System Bibliographic Services, which is operated by the Jet
Propulsion Laboratory, California Institute of Technology, under contract with the
National Aeronautics and Space Administration. S.C. acknowledges the financial
support of the Ministero della Ricerca Scientifica e dell'Universita' PRIN MIUR
2007: \lq{Multiple stellar populations in globular clusters}\rq\ (PI: G. Piotto),
ASI grant ASI-INAF I/016/07/0, and the Italian Theoretical Virtual Observatory
Project (PI: F. Pasian).

\end{document}

%% file: tab01.tex
\begin{deluxetable}{l|cccccc}
\tabletypesize{\scriptsize}
\tablewidth{0pt}
\tablecaption{Empirical estimates for the HB and RGB bump brightness, the \vhbbstar and the $R_{bump}$ parameters. \label{tab:tab1}}
\tablehead{\colhead{galaxy} & \colhead{$V^*_{bump}$} & \colhead{$V^*_{HB}$} & \colhead{\vhbbstar}  & \colhead{$R_{bump}$} }
\startdata
Cetus            &    $24.63\pm0.02$  &   $25.01\pm0.06$     &   $-0.38\pm0.06$  & $0.44\pm0.04$  \\
IC1613           &    $24.54\pm0.02$  &                      &   $-0.58\pm0.09$  & $0.50\pm0.03$  \\
                 &    $24.65\pm0.02$  &   $25.12\pm0.09$     &   $-0.47\pm0.09$  & $0.48\pm0.03$  \\
LGS 3            &    $24.00\pm0.02$  &   $24.61\pm0.06$     &   $-0.61\pm0.06$  & $0.46\pm0.05$  \\   
Tucana           &    $24.61\pm0.02$  &   $25.37\pm0.04$     &   $-0.76\pm0.05$  & $0.45\pm0.05$  \\
                 &    $24.88\pm0.02$  &                      &   $-0.50\pm0.05$  & $0.46\pm0.05$  \\
\enddata
\end{deluxetable}

%% file: tab02.tex
\begin{deluxetable}{l|cccccc}
\tabletypesize{\scriptsize}
\tablewidth{0pt}
\tablecaption{As Table~1 but for the values estimated from the solution CMDs. For each galaxy, we also list the metallicity and age range  of the stellar populations mostly contributing to the RGB bump feature. \label{tab:tab2}}
\tablehead{
\colhead{galaxy}  & \colhead{Age (Gyr)} & \colhead{Metalicity (Z)} & \colhead{$M_{V^*,bump}$} & \colhead{$M_{V^*,HB}$} &   \colhead{\vhbbstar} & \colhead{$R_{bump}$} }
\startdata
Cetus            &    $ > $ 8&   [0.0003, 0.0006]  &  $-0.35\pm0.02$  &  0.41  &  $-0.76\pm0.02$   & $0.48\pm0.03$ \\
                 &    $ > $ 8&   [0.0006, 0.0010]  &  $-0.15\pm0.02$  &  0.43  &  $-0.58\pm0.02$   & $0.55\pm0.06$ \\
IC1613           &    $ > $ 5&   [0.0006, 0.0012]  &  $-0.19\pm0.02$  &  0.43  &  $-0.62\pm0.02$   & $0.48\pm0.03$ \\
                 &    $ > $ 5&   [0.0012, 0.0020]  &  $-0.12\pm0.02$  &  0.45  &  $-0.57\pm0.02$   & $0.54\pm0.04$ \\
LGS3             &    $ > $ 5&   [0.0003, 0.0006]  &  $-0.39\pm0.02$  &  0.41  &  $-0.80\pm0.02$   & $0.44\pm0.02$ \\	
Tucana           &    $ > $ 8&   [0.0001, 0.0005]  &  $-0.31\pm0.02$  &  0.41  &  $-0.72\pm0.02$   & $0.44\pm0.02$ \\
                 &    $ > $ 8&   [0.0005, 0.0010]  &  $-0.12\pm0.02$  &  0.45  &  $-0.57\pm0.02$   & $0.42\pm0.02$ \\
\enddata
\end{deluxetable}

%% file: tab03.tex
\begin{deluxetable}{lrr}
\tabletypesize{\scriptsize}
\tablewidth{0pt}
\tablecaption{Estimates of the \vhbb parameter and of the metallicity of the associated stellar populations for selected LG dwarfs from the literature. }
\tablehead{
\colhead{Galaxy}   &  \colhead{\vhbb} & \colhead{[M/H]}     } 
\startdata
Sculptor           & $-0.90\pm0.21 $\tablenotemark{a}  &   $-2.0\pm0.2$\tablenotemark{b}    \\ 
Sculptor           & $-0.35\pm0.21 $\tablenotemark{a}  &   $-1.30\pm0.15$\tablenotemark{b}  \\ 
Ursa Minor         & $-0.62\pm0.11 $\tablenotemark{c}  &   $-1.7\pm0.2$\tablenotemark{c}    \\
Sagittarius        & $ 0.22\pm0.10 $\tablenotemark{d}  &   $-0.45\pm0.15$\tablenotemark{d}  \\ 
Sextans            & $-0.45\pm0.06 $\tablenotemark{e}  &   $-1.63\pm0.20$\tablenotemark{f}  \\
\enddata
\tablenotetext{a}{\citealt{majewski99};}
\tablenotetext{b}{\citealt{kirby09};}
\tablenotetext{c}{\citealt{bellazzini02};}
\tablenotetext{d}{\citealt{monaco02};}
\tablenotetext{e}{\citealt{lee03};}
\tablenotetext{f}{\citealt{lee09}.}
\label{tab:tab3}
\end{deluxetable}